\documentclass[,twocolumn,,trackchanges]{aastex631}
\usepackage{amssymb}
\usepackage{amsmath}
\usepackage{epstopdf}
\usepackage{natbib}
\usepackage{xcolor}
\usepackage[T1]{fontenc}
\usepackage{url}
\usepackage{epsfig}
\usepackage[mathscr]{euscript}
\usepackage{threeparttable}

\begin{document}

\linespread{1.12}\selectfont
\title{Propagation of Precessing Jet in Envelope of Tidal Disruption Events}

\correspondingauthor{Wei-Hua Lei}
\email{leiwh@hust.edu.cn}

\author{Hao-Yu Yuan}
\affiliation{Department of Astronomy, School of Physics, Huazhong University of Science and Technology, Luoyu Road 1037, Wuhan, 430074, China}

\author{Hong-Zhou Wu}
\affiliation{Department of Astronomy, School of Physics, Huazhong University of Science and Technology, Luoyu Road 1037, Wuhan, 430074, China}

\author[0000-0003-3440-1526]{Wei-Hua Lei}
\affiliation{Department of Astronomy, School of Physics, Huazhong University of Science and Technology, Luoyu Road 1037, Wuhan, 430074, China}


\begin{abstract}
It is likely that the disk of a tidal disruption event (TDE) is misaligned with respect to the equatorial plane of the spinning supermassive black hole (SMBH), since the initial stellar orbit before disruption is most likely has an inclined orbital plane. Such misaligned disk undergoes Lense-Thirring precession around the SMBH spin axis, leading to a precessing jet if launched in the vicinity of the SMBH and aligned with the disk angular momentum. The bound debris can also build a thick envelope which powers optical emission. In this work, we study the propagation of the precessing jet in the TDE envelope. We adopt a ``zero-Bernoulli accretion'' (ZEBRA) envelope model. A episodic jet will be observed if the line of sight is just at the envelope pole direction and $\theta_{\rm LT}=\theta_{\rm env}$, since the jet can freely escape from this low density rotation funnel, where $\theta_{\rm LT}$ and $\theta_{\rm env}$ are the jet precessing angle and the angle between the envelope polar axis and the SMBH spin axis, respectively. The jet will be choked at other directions. For $\theta_{\rm LT} < \theta_{\rm env}$, the jets can also break out of the envelope for very small precession angle $\theta_{\rm LT}$ or if the jet is aligned with SMBH spin. If the jet is choked within the envelope, the radiation produced during cocoon shock breakout will imprint characteristic signatures on the X-ray emission, such as low-amplitude fluctuation in the light curve.


\end{abstract}

\keywords{\href{http://astrothesaurus.org/uat/1696}{Tidal disruption (1696)}}


\section{Introduction}\label{Intro}
A star would be disrupted by a supermassive black hole (SMBH) when it comes too close to the central SMBH. In such events, namely tidal disruption events (TDEs), the stellar debris are eventually accreted by the SMBH, producing bright multi-band emission \citep{Hills(1975), Rees(1988)}. The initial stellar orbit is most likely misaligned with the equatorial plane of the spinning SMBH. The tilted disk formed after the TDE would undergo Lense-Thirring precssion around the SMBH spin axis \citep[e.g.,][]{Papaloizou-Lin(1995), Fragile-et-al.(2007), Stone2012, Lei2013, WangJZ2014,Franchini2016,Li2025}. The $\sim 15$ day quasi-periodic modulations found in the high-cadence X-ray monitoring observations of TDE candidate AT2020ocn, suggest the Lense-Thirring precession of a newly formed accretion disk \citep{Pasham2024}.


Relativistic jets have been detected in some TDEs, which generate non-thermal hard X-ray emission and radio afterglow, e.g., Swift J1644+57 \citep[hereafter J1644,][]{Bloom-et-al.(2011), Burrows-et-al.(2011), Levan-et-al.(2011)}, Swift J2058.4+0516 \citep[hereafter J2058,][]{Pasham-et-al.(2015)}, Swift J1112.28238 \citep[hereafter J1112,][]{Brown-et-al.(2015)} and AT 2022cmc \citep{Andreoni-et-al.(2022), Yao-et-al.(2024), Zhou2024}. These finds had stimulated a great interest in studying jets launched from TDEs. Quasi-periodic oscillations in the light curves have been observed in jetted TDE J1644 \citep{Burrows-et-al.(2011),Saxton2012}, suggesting a precessing jet due to Lense-Thirring effects would come into play \citep{Stone2012,Lei2013}.

For jetted TDEs, a question emerges: why would only a small fraction of TDEs successfully launch jets? One factor is the narrow beaming cone of the jets, and our line of sight is usually away from such jet cones. A recent study on IGR J12580+0134 revealed the first TDE with an off-axis relativistic jet \citep{Lei2016, YuanQ2016}. \citet{Dai-et-al.(2018)} proposed a unified model based on 3D general relativistic radiation magneto-hydrodynamis (GRRMHD) simulation that can explain both the optical and X-ray TDEs. The observed emission type mainly depends on the viewing angle of the observer with respect to the disk orientation. X-ray emissions will be observed from the optically thin funnel surrounded by winds when viewed face-on. For edge-on observations, only optical/UV emissions can be detected due to the heavily reprocessing of X-ray emissions in the chick wind and disk \citep{Dai-et-al.(2018),Thomsen2022, Chen-et-al.(2025)}. However, even after the correction of relativistic beaming of jet and view angle effects, the fraction of jetted TDEs ($10^{-3}$ to $10^{-2}$ of all TDEs) is still much smaller than the fraction of active galactic nuclei with jets ($\sim 10$ per cent) \citep{Teboul-Metzger(2023), Lu-et-al.(2024)}. This remains an open question for jetted TDEs. 


\citet{Teboul-Metzger(2023)} and \citet{Lu-et-al.(2024)} highlight another critical factor: jet precession. \citet{Teboul-Metzger(2023)} pointed out that powerful jets can successfully escape from disk wind, while less powerful jet can only escape after alignment with the SMBH spin. \citet{Lu-et-al.(2024)} showed that precessing jets can break out of the disk wind if the misalignment angle is less than a few times of the jet opening angle. For large misalignment angles, the jets are initially choked by the disk wind and may only break out later when the disk eventually aligns with the SMBH spin. The very small event rate of jetted TDEs is thus explained by the conditions (alignment, orbital penetration factor, jet efficiency) \citep{Teboul-Metzger(2023),Lu-et-al.(2024)}. In these pictures, the long hydrodynamic alignments give rise to late jet escape and delayed radio flares as seen in many optically bright TDEs \citep{Teboul-Metzger(2023),Lu-et-al.(2024)}. 

Besides the disk wind scenarios, envelope models are also proposed to explain the bright optical/UV emissions in TDEs \citep{Loeb-Ulmer(1997), Coughlin-Begelman(2014)}. \citet{Loeb-Ulmer(1997)} developed a steady, optically-thick, spherical envelope surrounding an inner accretion disk. The disk X-ray emissions are reprocessed into optical/UV emissions by the envelope. \citet{WangXY2016} studied the TDE jet propagation in an extended, optically thick envelope. Neutrino emission is predicted to be produced in the choked jets of TDEs \citep{WangXY2016, Zheng2023}. Considering the angular momentum transportation between disk and envelope, \citet{Coughlin-Begelman(2014)} proposed a model (called ``Zero Bernoulli accretion flow'', or ZEBRA) in which the infalling gas traps accretion energy and inflates into a quasi-spherical structure with zero Bernoulli parameter. In this ZEBRA envelope model, most of the accretion energy escapes through the poles in the form of powerful jets \citep{Coughlin-Begelman(2014)}. Recently, general relativistic hydrodynamical simulations revealed the production of the optically thick envelope with an inner region, confirming envelopes as the likely solution to the puzzle of TDE optical/UV emissions \citep{Price2024}. These studies motivate us to investigate the propagation of a precessing jet in the ZEBRA envelope.

Since jets can escape directly from the poles of ZEBRA envelope, and the poles are mostly likely misaligned with the SMBH spin axis (the ZEBRA would keep the initial stellar angular momentum), a new factor determining the escaping of precessing jet in our model is the ZEBRA inclination angle (assumed to be the initial stellar orbit inclination). This is the main difference with the previous work.

The paper is organized as follows: We review the ZEBRA envelope model in Sec. \ref{ZEBRA} and consider the propagation of a precessing jet in an inclined envelope in Sec. \ref{precessing}. The results of the jet breakout conditions are presented in Sec. \ref{sec: results}. In Sec. \ref{sec: discussion}, we compare the model with previous work and the TDEs observations, and draw conclusions in Sec.\ref{conclusion}.



\section{Model Descriptions}\label{model}
In this section, we will summarize the ZEBRA envelope model in Sec. \ref{ZEBRA}. In Sec. \ref{precessing}, we delineate both the dynamical processes and morphological characteristics of the precession jet in an inclined ZEBRA envelope. 

\subsection{ZEBRA Envelope}\label{ZEBRA}
The ZEBRA model is used to describe a inflated envelope with a Bernoulli parameter of zero, which is developed based on the work of \citet{Loeb-Ulmer(1997)}.
Compared with \citet{Loeb-Ulmer(1997)}, \citet{Coughlin-Begelman(2014)} made the following modifications and improvements:
\begin{itemize}
    \item The angular momentum within the envelope is transferred outward, causing the inner boundary of the envelope to shrink.    
    \item The shrinkage of the inner boundary further leads to super-Eddington accretion, and the energy generated by this process is absorbed by the envelope.
    \item The observed super-Eddington luminosity and relativistic jets indicate that SMBH accretion is not self-regulated. \citet{Coughlin-Begelman(2014)} assumed that the structure of the flow is regulated by its ratio of angular momentum to mass, which is quite sub-Keplerian between the vicinity of the SMBH and the photosphere. At this point, the accretion loses its self-regulation capability, and the energy produced by accretion is absorbed by the envelope material, expanding outward until it reaches the position of weakly bound. Besides, the envelope simultaneously forms low-density polar funnels at its two poles.    
    \item The envelope can not fully absorb the energy released by accretion, and the excess energy is released in the form of relativistic jets through the polar funnels.
\end{itemize}
The remaining content of this section will briefly introduce the structure and evolution of the envelope.


Neglecting the relativistic corrections, the Bernoulli parameter of the flow can be expressed as, $B=\Omega^2r^2/2-GM_{\bullet}/r+H$, where $\Omega$ is the angular velocity, $H$ is the enthalpy, $M_{\bullet}$ is the mass of SMBH and $r$ is the distance from the SMBH \citep{Narayan-Yi(1994)}. The momentum equations and the
Bernoulli equation are expressed as follow within the spherical coordinates,
\begin{eqnarray}
\label{ZEBRA euqations}
\nonumber\frac{1}{\rho}\frac{\partial p}{\partial r}=-\frac{GM_{\bullet}}{r^2}+\frac{l^2\csc ^2\theta}{r^3},\\ 
\frac{1}{\rho}\frac{\partial p}{\partial \theta}=\frac{l^2\cot \theta\csc ^2\theta}{r^2},\\ \nonumber
-\frac{GM_{\bullet}}{r}+\frac{l^2\csc^2\theta}{2r^2}+\frac{\gamma}{\gamma-1}\frac{p}{\rho}=0, \nonumber
\end{eqnarray}
where $l$ is the specific angular momentum of the gas, $p$ is the gas pressure, $\theta$ is the polar angle and $\gamma \approx 3/4$ is the adiabatic index of the gas for the radiation pressure domination. 
\citet{Blandford-Begelman(2004)} derived a self-similar solution for the density, pressure and specific angular momentum,
\begin{eqnarray}
\label{ZEBRA solution}
\rho(r,\theta)=\rho_0(r/r_0)^{-q} (\sin^2\theta)^\alpha,\\
p(r,\theta)=\beta \frac{GM_{\bullet}\rho_0}{r}(r/r_0)^{-q} (\sin^2\theta)^\alpha,\\
l^2(r,\theta)=aGM_{\bullet}r\sin^2\theta,
\end{eqnarray}
where $q=3/2-n$, $\alpha=\frac{1-q(\gamma-1)}{\gamma-1}$, $\beta=\frac{\gamma-1}{1+\gamma-q(\gamma-1)}$, and $a=\frac{2-2q(\gamma-1)}{1+\gamma-q(\gamma-1)}$.
$r_0\sim \chi r_{\rm s}$ is the inner radius of the envelope. Here, $r_{\rm s}=2GM_\bullet/c^2$ is the Schwarzschild radius, and $\chi$ is order of a few. $\rho_0$ is the density at this radius. The parameter $n$ is defined by Blandford and Begelman so that the accretion rate $\propto r^n$. 


One can find that $n$, and therefore $q$, determines the structure and evolution of the ZEBRA envelope. The total mass $M=\int_{r_0}^{r_{\rm out}}\rho dV$ and angular momentum $L=\int_{r_0}^{r_{\rm out}}\rho ldV$ of the envelope are,
\begin{equation}
\label{Eq:total mass}
M=\frac{2\pi^{3/2}\rho_0}{r_0^{-q}}\frac{\Gamma(\alpha+1)}{\Gamma(\alpha+3/2)}\frac{r_{\rm out}^{-q+3}}{-q+3},
\end{equation}

\begin{equation}
\label{Eq:total angular moment}
L=\frac{2\pi^{3/2}\rho_0}{r_0^{-q}\sqrt{aGM_{\bullet}}}\frac{\Gamma(\alpha+3/2)}{\Gamma(\alpha+2)}\frac{r_{\rm out}^{-q+7/2}}{-q+7/2},
\end{equation}
where $\Gamma(x)$ represents the $\Gamma$-functions and $r_{\rm out}\gg r_0$ is the outer radius of the envelope.

The ZEBRA model assumes that the energy absorbed within the envelope is transported outward via convection. At the outer boundary of the envelope, the radiative luminosity is maintained at the Eddington limit, meaning that the convective luminosity ($\int ypc_{\rm s}dS$) at the outer boundary equals the Eddington luminosity ($4\pi GcM_{\bullet}/\kappa)$, where $y$ is the efficiency of advection, $c_{\rm s}$ is the local sound velocity, $\kappa\approx 0.34\rm ~cm^2~g^{-1}$ is the opacity and $dS$ is the surface of the envelope. One can derive that,
\begin{equation}
\label{Eq:r_out}
r_{\rm out}^{-q+1/2}=\frac{2c}{\kappa \sqrt{\pi}} \frac{\Gamma(\alpha+3/2)}{\Gamma(\alpha+1)}\frac{r_0^{-q}}{\rho_0 y\beta\sqrt{aGM_{\bullet}}},
\end{equation}
combining the Eqs. (\ref{Eq:total mass}), (\ref{Eq:total angular moment}) and (\ref{Eq:r_out}), one can yield,
\begin{eqnarray}
\label{Eq:q solution}
\nonumber\left(\frac{y\kappa}{4\pi c}\right)^{1/6}\frac{M\sqrt{GM_{\bullet}}}{L^{5/6}}\\
=\frac{\Gamma(\alpha+1)^{5/6}\Gamma(\alpha+2)^{5/6}}{\beta^{1/6}a^{1/2}\Gamma(\alpha+3/2)^{5/3}}\frac{(7/2-q)^{5/6}}{3-q}.
\end{eqnarray}
Therefore, knowing the time-evolution of the envelope mass, we can determine the evolution of $q$ and ultimately derive the structural evolution of the envelope. The evolution of the total mass of the envelope is $dM/dt=\dot{M}_{\rm fb}-\dot{M}_{\rm acc}$, where $\dot{M}_{\rm fb}$ is the rate at which material returns to pericenter and $\dot{M}_{\rm acc}$ is the mass accretion rate onto SMBH. The infall velocity ($v_{r_0}$) of material at the inner boundary during accretion onto the black hole is not the free-fall velocity due to the influence of radiation pressure, and we adopt $v_{r_0}=\delta\sqrt{GM_{\bullet}/r_0}$ with $\delta$ as a parameter less than one, and $\dot{M}_{\rm acc}\approx 4\pi r_0^2\rho_0v_{r_0}$. $\dot{M}_{\rm fb}$ is determined following the method described in 
\citet{Lodato-et-al.(2009)}. 


The luminosity of the jet ($L_{\rm j}$) and the effective temperature at the outer boundary of the envelope ($T_{\rm env}$) are,
\begin{equation}
\label{Eq:jet luminosity}
L_{\rm j}=\epsilon \dot{M}_{\rm acc}c^2,
\end{equation}
where $\epsilon$ is the efficiency of accretion to radiation.
\begin{equation}
\label{Eq:envelope temperature}
T_{\rm env}=\left(\frac{GcM_{\bullet}m_{\rm p}}{\sigma_{\rm T}\sigma_{\rm SB}r_{\rm out}^2}\right)^{1/4},
\end{equation}
where $\sigma_{\rm SB}$ is the Stefan–Boltzmann constant and $\sigma_{\rm T}$ is the Thomson scattering cross-section.

\begin{figure*}
\centering
\includegraphics [angle=0,scale=0.25] {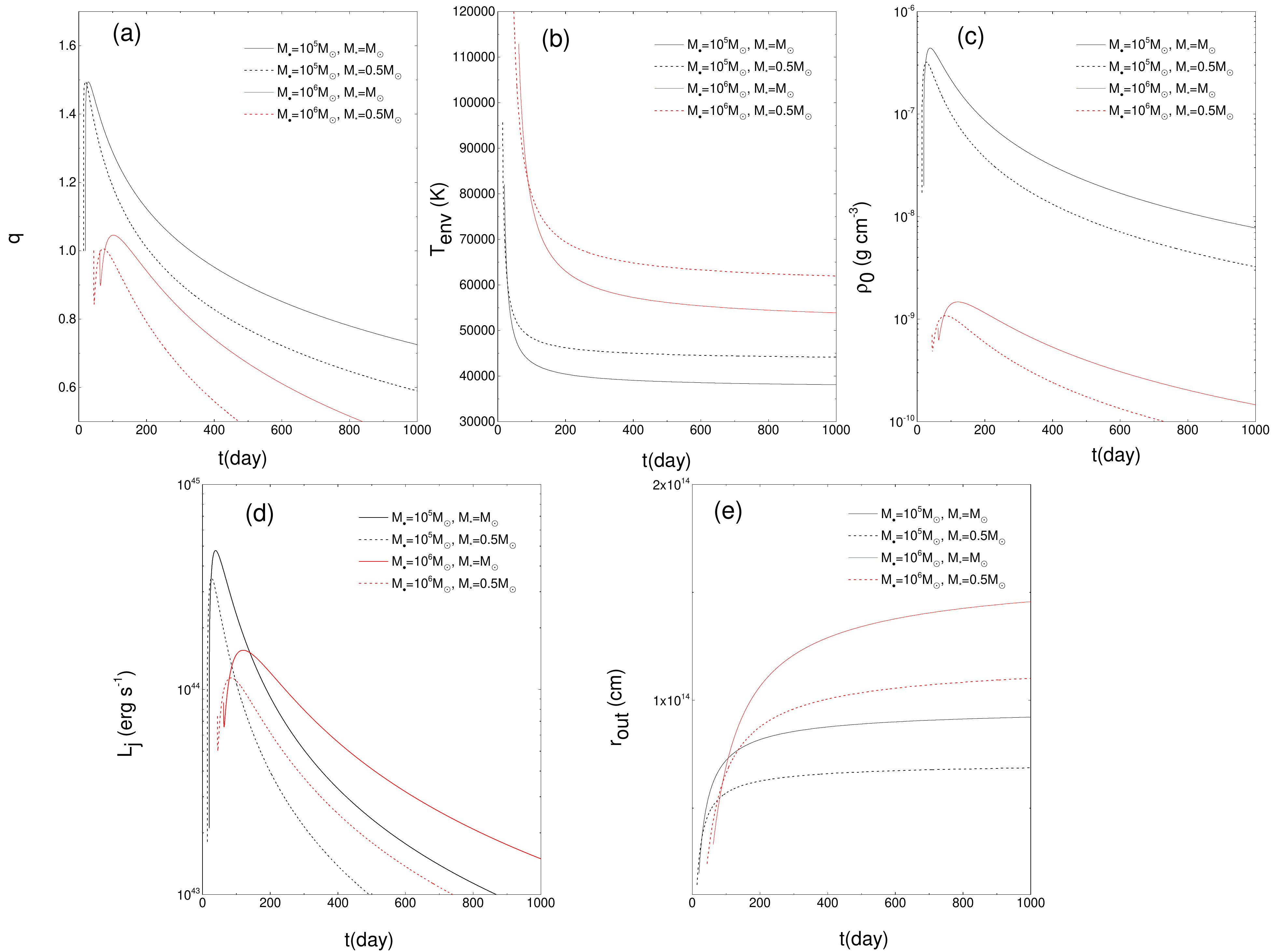}
\caption{The solution of q with time (Panel a) and the characteristics of the ZEBRA envelope evolve as a function of time, i.e., the effective temperature ($T_{\rm env}$, panel b), the density at the inner radius ($\rho_0$, panel c), the jet luminosity ($L_{\rm j}$, panel d) and the outer radius ($r_{\rm out}$, panel e).  In the calculation, we adopt the initial value of $q$ as $q_0=1$, for the mass disrupted star $M_*= M_\odot$ (solid line) and $M_*= 0.5 M_\odot$ (dashed line) with $M_{\bullet}=10^5M_\odot$ (black line) and $10^6M_\odot$ (red line). And we use the accretion efficiency $\epsilon=0.01$, $\chi =5$, $y=0.5$ and $\delta=0.05$.
}
\label{fig:ZEBRA_character}
\end{figure*}

Fig. \ref{fig:ZEBRA_character} illustrates the time evolution of the $q$ and the global properties of the ZEBRA envelope. We consider the disruption of star with mass $M_*=M_\odot$ (solid lines) and $M_*=0.5M_\odot$ (dashed lines) by SMBH with mass $M_{\bullet}=10^5M_\odot$ (black lines) and $M_{\bullet}=10^6M_\odot$ (red lines). We show the evolution of $q$ in the panel (a). One can find that $M_{\bullet}$ and $M_*$ have significant effects for the whole evolution of $q$. 
In the panels (b)-(e), we also display the evolution of $T_{\rm env}$, $\rho_0$, $L_{\rm j}$ and $r_{\rm out}$ for different $M_*$ and $M_{\bullet}$. 
In addition, one should note that the accretion rate of SMBH would be lower than the Eddington accretion rate for $M_\bullet \sim 10^{7}M_\odot$\citep{Coughlin-Begelman(2014)}, and the ZEBRA envelope may not be formed. Therefore we do not consider such case in this work. 

The ZEBRA model described above does not take into account the fact that, for majority cases, the orbital angular momentum of the disrupted star would be misaligned with the SMBH spin. Therefore, the envelope is most likely inclined with respect to the SMBH spin axis due to such misalignment. We consider an inclined ZEBRA envelope, and rewrite the density, pressure, and specific angular momentum equations as follows,
\begin{eqnarray}
\label{ZEBRA solution rewrite}
\rho(r,\Delta\theta_{\rm env})=\rho_0(r/r_0)^{-q} (\sin^2\Delta\theta_{\rm env})^\alpha\\
p(r,\Delta\theta_{\rm env})=\beta \frac{GM_{\bullet}\rho_0}{r}(r/r_0)^{-q} (\sin^2\Delta\theta_{\rm env})^\alpha\\
l^2(r,\Delta\theta_{\rm env})=aGM_{\bullet}r\sin^2\Delta\theta_{\rm env}
\end{eqnarray}
where $\Delta\theta_{\rm env}$ is the position angle of the fluid element with respect to polar axis of the envelope. We adopt the azimuthal angle of the polar axis of the envelope as $\phi_{\rm env}=0$, one can derive that the cosine of $\Delta\theta_{\rm env}$ as $\cos \Delta\theta_{\rm env}=\sin\theta_{\rm env}\sin\theta\cos\phi+\cos\theta\cos\theta_{\rm env}$, where $\theta_{\rm env}$ is the angle between the envelope polar axis and the SMBH spin axis. 

This change will not significantly alter the results, since the solutions mainly depend on the total mass and angular momentum of the envelope. Therefore, for the inclined ZEBRA envelope, we adopt the same structure and evolution behaviors as presented in Fig. \ref{fig:ZEBRA_character}. The typical values for the parameters we adopted are $\epsilon=0.01$, $\chi =5$, $y=0.5$ and $\delta=0.05$ \citep{Coughlin-Begelman(2014)}.

\subsection{Precessing Jet}\label{precessing}

\begin{figure*}
\centering
\includegraphics [angle=0,scale=0.6] {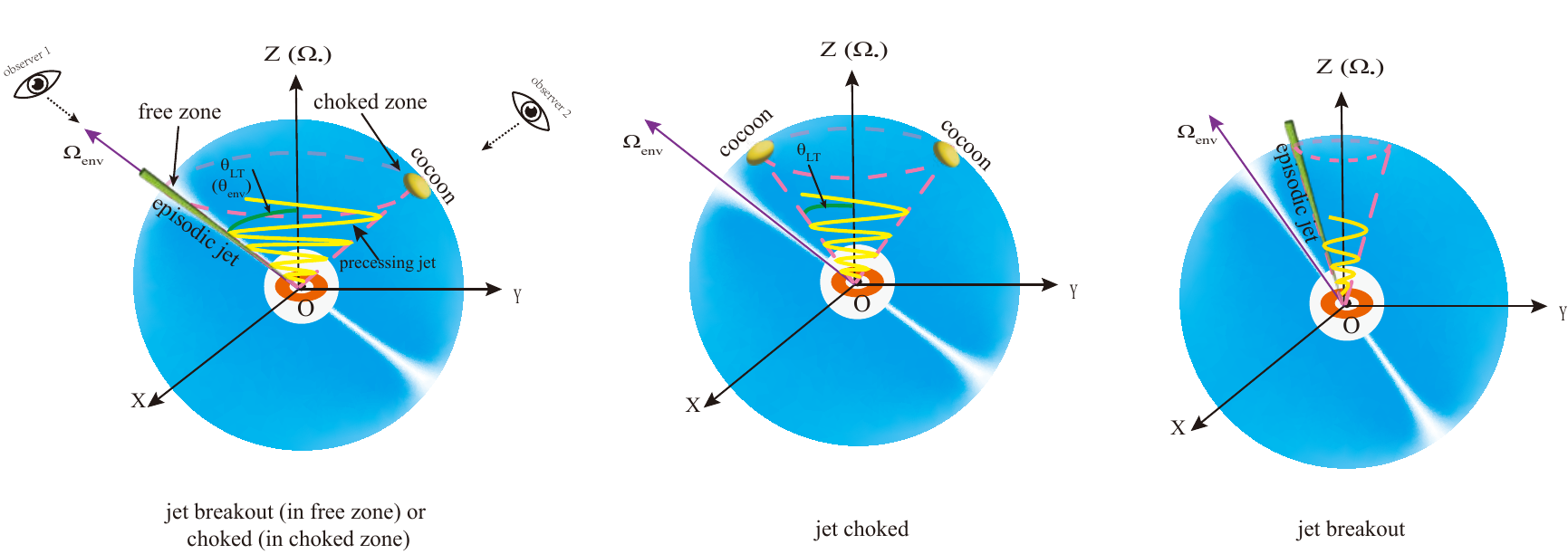}
\caption{A schematic picture of the precession jet in an inclined ZEBRA envelope, the central orange ring represents the inner accretion disk within the ZEBRA envelope. The jet precess with angle $\theta_{\rm LT}$. The angle between the SMBH spin axis ($\Omega_{\bullet}$) and the envelope's rotation axis ($\Omega_{\rm env}$), i.e., the envelope inclination, is $\theta_{\rm env}$. 
Left panel: 
$\theta_{\rm LT}=\theta_{\rm env}$. In the low density polar regions, i.e, ``free zone'', the relativistic jet can escape freely when it precess to this direction. The jet is choked at other regions. 
Middle and right panels: $\theta_{\rm LT}<\theta_{\rm env}$. The jet will be choked in all directions if $\theta_{\rm LT}$ is large (middle panel), or break out of envelope if $\theta_{\rm LT}$ is small (right panel).
}
\label{fig:ZEBRA_model}
\end{figure*}

We consider an inner disk locating at the center of the envelope \citep{Loeb-Ulmer(1997)}. The envelop is much large in size, and we assume that only the inner disk undergoes significant Lense-Thirring precession.

Now, we introduce a precessing jet (with opening angle $\theta_{\rm j}$) launching from the center inner disk of the inclined ZEBRA envelope. The inner accretion disk undergoes Lense-Thirring precession with an precessing angle $\theta_{\rm LT} \leqslant \theta_{\rm env}$. We assume that the jet is normal to the inner disk and precess with the same period $P_{\rm LT}$. We show the schematic picture of our model in Fig. \ref{fig:ZEBRA_model}. 

\begin{figure}
\centering
\includegraphics [angle=0,scale=0.35] {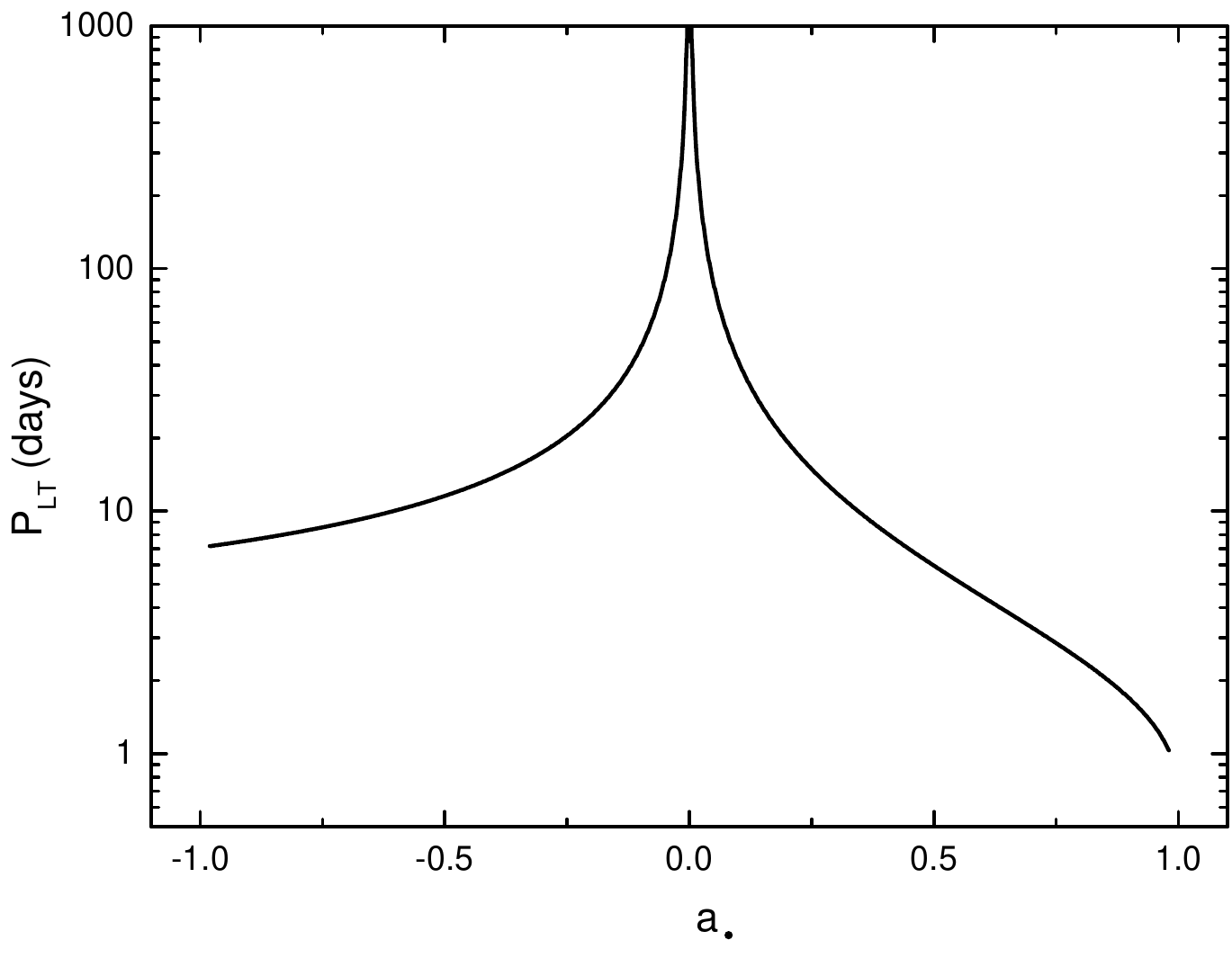}
\caption{The precessing period $P_{\rm LT}$ as a function of SMBH spin $a_\bullet$ for $M_{\bullet}\sim 10^6M_\odot$ and $M_*\sim M_\odot$.
}
\label{fig:P_LT_spin}
\end{figure}

Due to the super-Eddington accretion, the inner disk is expected to be thick \citep{Ulmer(1997)}, and the surface density of the disk follows $\Sigma\propto r^{\rm -\zeta}$.
The inner radius of the disk $r_{\rm d,in}$ is the innermost stable circular orbit (ISCO) and the outer radius $r_{\rm d,out}$ is the circularization radius of the TDE debris.
$P_{\rm LT}$ depends on the SMBH mass $M_{\bullet}$ and spin $a_{\bullet}$, the mass of disrupted star $M_*$ and the boundaries of the precessing disk. According to the method of \citet{Franchini2016} and \citet{Stone2012}, 

\begin{eqnarray}
\label{Eq:Period}
P_{\rm LT}=\frac{8\pi G M_\bullet (1+2)\zeta}{c^3(5-2\zeta)} \frac{ \xi_{\rm out}^{5/2-\zeta} \xi_{\rm in}^{1/2+\zeta}\left[1-(\frac{\xi_{\rm in}}{\xi_{\rm out}})^{5/2-\zeta}\right]}{a_\bullet \left[1-(\frac{\xi_{\rm in}}{\xi_{\rm out}})^{1/2+\zeta}\right]}, \nonumber \\
\end{eqnarray}
where $\xi_{\rm in} = r_{\rm d,in}/r_{\rm s}$ and $\xi_{\rm out} = r_{\rm d,out}/r_{\rm s}$. We calculate the $P_{\rm LT}$ evolution with SMBH spin $a_\bullet$ for $M_{\bullet}\sim 10^6M_\odot$, $M_*\sim M_\odot$ and $\zeta=0.6$, as shown in Fig.\ref{fig:P_LT_spin}.
One can get $P_{\rm LT}\sim 40~$days for $a_{\bullet}\sim 0.1$, $P_{\rm LT}\sim 6~$days for $a_{\bullet}\sim 0.5$ and $P_{\rm LT}\sim 1~$day for $a_{\bullet}\sim 1$.

For an observer with line of sight at a polar angle $\theta_{\rm obs}\in (\theta_{\rm LT}-\theta_{\rm j},\theta_{\rm LT}+\theta_{\rm j})$, the jet sweeps the line of sight periodically. The jet's observable duration in one precessing is $t_{\rm on}$, and the duty cycle of the episodic jet can be defined as,
\begin{equation}
\label{Eq:duty cycle}
\xi_{\rm duty}=\frac{t_{\rm on}}{P_{\rm LT}}.
\end{equation}

Assuming the observer’s azimuthal angle $\phi_{\rm obs}\in (0,2\pi)$, we can explore the conditions for successful jet breakout in this direction.

The jet duration is $T_{\rm j}=\xi_{\rm duty}P_{\rm LT}$ and the width of the shell is $\Delta_{\rm j}\approx T_{\rm j} c$. The jet would be choked if its tail catches up to the head insider the ZEBRA envelope photosphere $r_{\rm ph}$ (defined by $\int_{r_{\rm ph}}^{r_{\rm out}}\kappa\rho dr=1$). Assuming the catch-up time $t_{\rm c}$ in the lab frame and the catch-up radius $r_{\rm c}$, one finds $(v_{\rm t}-v_{\rm h})t_{\rm c} =\Delta_{\rm j}$, where $v_{\rm t}\approx c$ and $v_{\rm h}\approx c\sqrt{1-1/\Gamma_{\rm j}^2}$ are the tail and head velocities of the jet, and $\Delta_{\rm j}\approx cT_{\rm j}$ is the length of the jet. The jet catch-up time is $t_{\rm c}=\Delta_{\rm j}/(v_{\rm t}-v_{\rm h})=T_{\rm j}/(\beta_{\rm t}-\beta_{\rm h})\approx T_{\rm j}/(1-\beta_{\rm h})\approx 2\Gamma_{\rm j}^2 T_{\rm j}$, and $\Gamma_{\rm j}$ is the jet bulk Lorentz factor which is determined by the jet dynamics, $\beta_{\rm t}=v_{\rm t}/c$ and $\beta_{\rm h}=v_{\rm h}/c$. The catch-up radius is given by $r_{\rm c}=v_{\rm t} t_{\rm c}$. The jet choke condition is therefore $r_{\rm c}<r_{\rm ph}$.




During the jet propagation in ambient medium, e.g., the ZEBRA envelope medium, two shocks are generated, the forward and reverse shocks (FS and RS). The shocks and the jet head can divide the system into four region: (1) unshocked ZEBRA medium, (2) shocked ZEBRA medium, (3) shocked jet shell material and (4) unshocked jet shell material.

In the observer frame, the radius of the  deceleration $r_{\rm dec}$
is a radius over which the accumulated ZEBRA medium mass is $1/\Gamma_0$ of the jet isotropic mass \citep{Zhang-et-al.(2003), Meszaros-et-al.(1998), Kobayashi(2000)}, 
which can be defined as $\int_{r_0}^{r_{\rm dec}}4\pi r^2\Gamma_0\rho dr=E_{\rm iso}/\Gamma_0c^2$, 
and the deceleration time $t_{\rm dec}=r_{\rm dec}/2\Gamma_0^2c$, where $\Gamma_0$ is the initial Lorentz factor of the jet and $E_{\rm iso}=L_{\rm j}T_{\rm j}/(1-\cos\theta_{\rm j}$). 
The Lorentz factors of region 2 and 3 are \citep{Sari-Piran(1995)}, 
\begin{equation}
\label{eq: Lorentz_factors}
\gamma_2=\gamma_3\approx \left\{
\begin{aligned}
&\Gamma_0, &t_{\rm dec}<T_{\rm j}~ \text{(thin shell regime)} \\
&\Gamma_0^{1/2}f^{1/4}/\sqrt{2}, &t_{\rm dec}>T_{\rm j}~ \text{(thick shell regime)}
\end{aligned}
\right. 
\end{equation}
where $f=n_4/n_1$ is the number density ratio between region 4 and 1, and $n_4=E_{\rm iso}/(4\pi m_{\rm p}c^2r^2\Gamma_0^2\Delta_{\rm j})$ and $n_1$ are the number density of the region 4 and 1, and $\Gamma_{\rm j}\sim \gamma_2$. The choke time will be $t_{\rm c}=t_{\rm dec}$ for the thin shell regime, and $t_{\rm c}=T_{\rm j}$ for the thick shell regime.

\section{Results}\label{sec: results}



\begin{figure*}
\centering
\includegraphics [angle=0,scale=0.35] {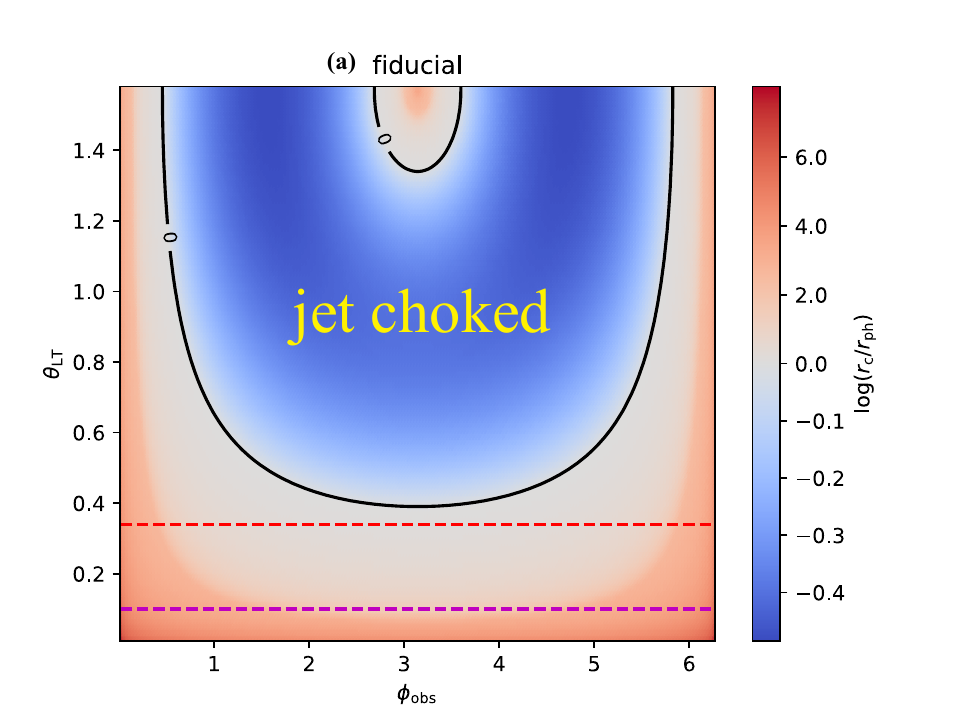}
\includegraphics [angle=0,scale=0.35] {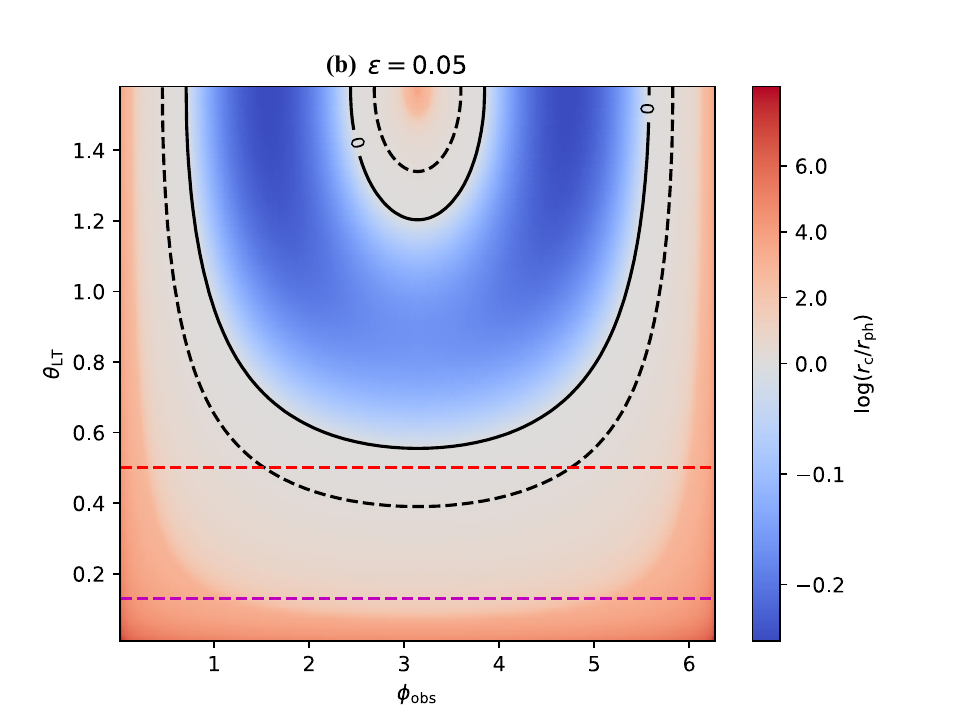}
\includegraphics [angle=0,scale=0.35] {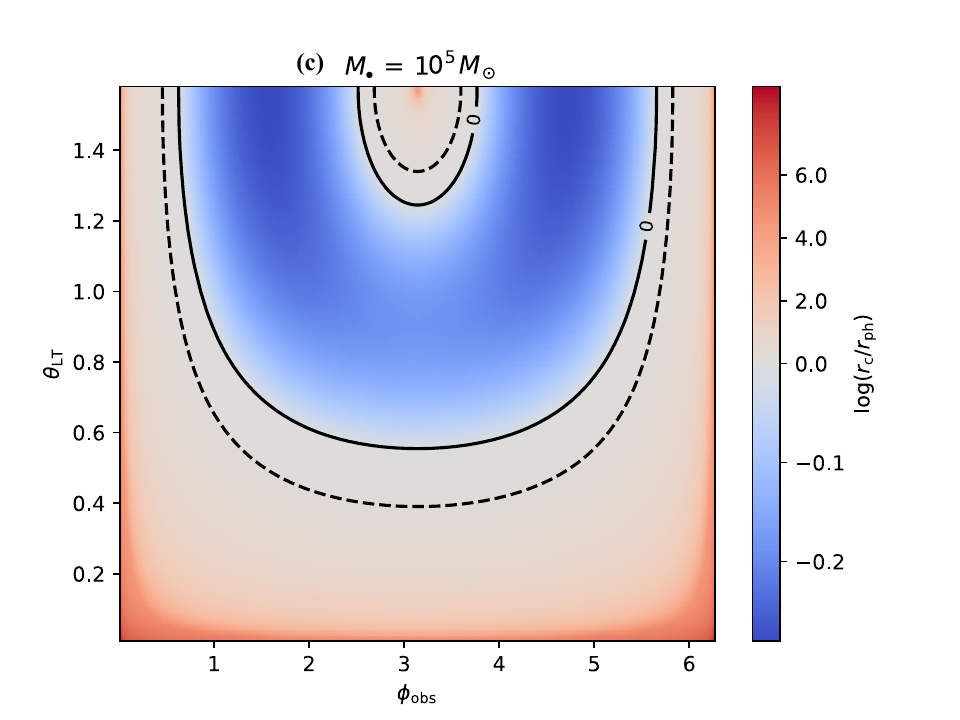}
\includegraphics [angle=0,scale=0.35] {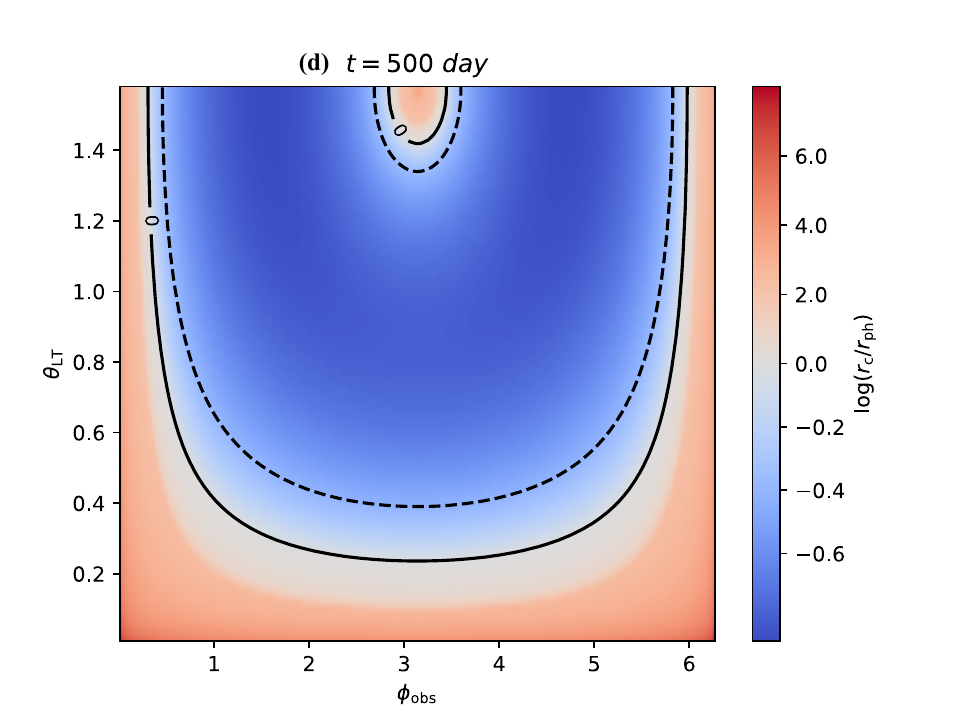}
\includegraphics [angle=0,scale=0.35] {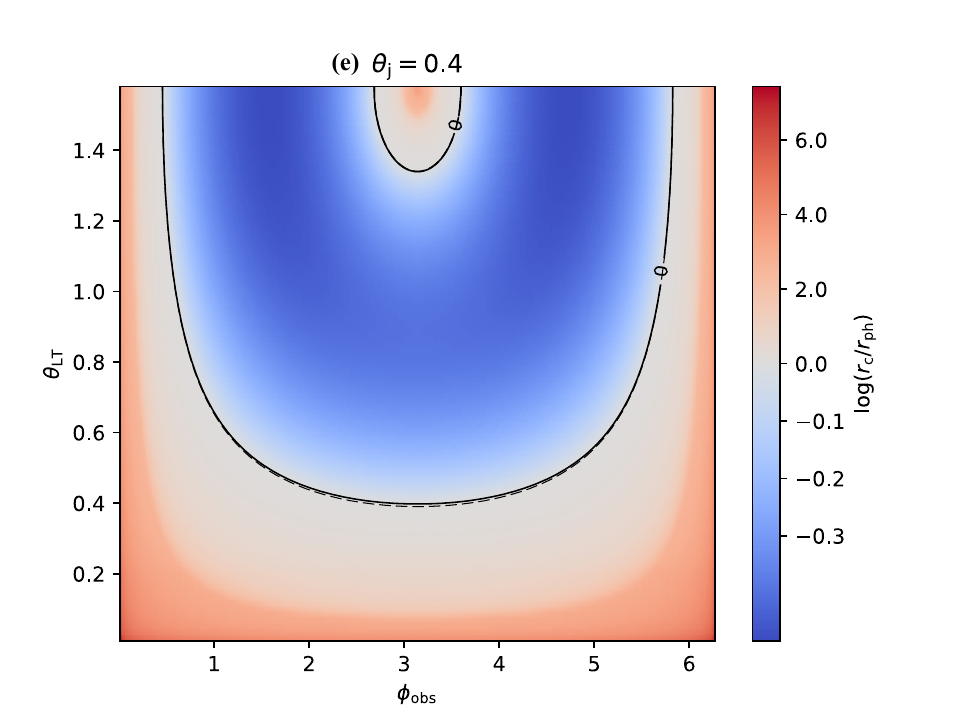}
\includegraphics [angle=0,scale=0.35] {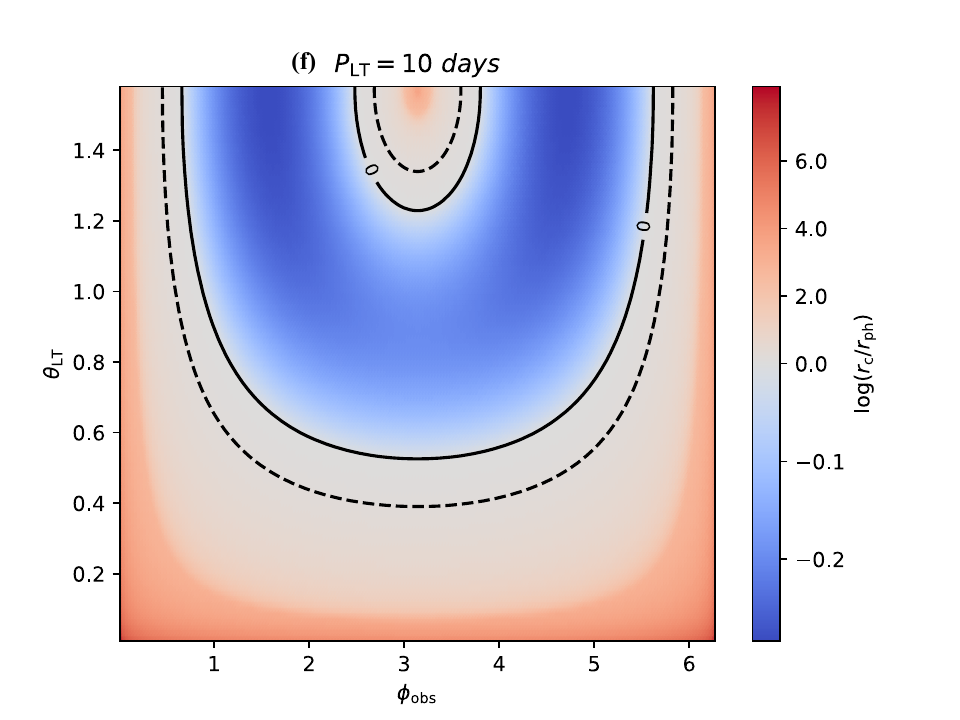}
\includegraphics [angle=0,scale=0.35] {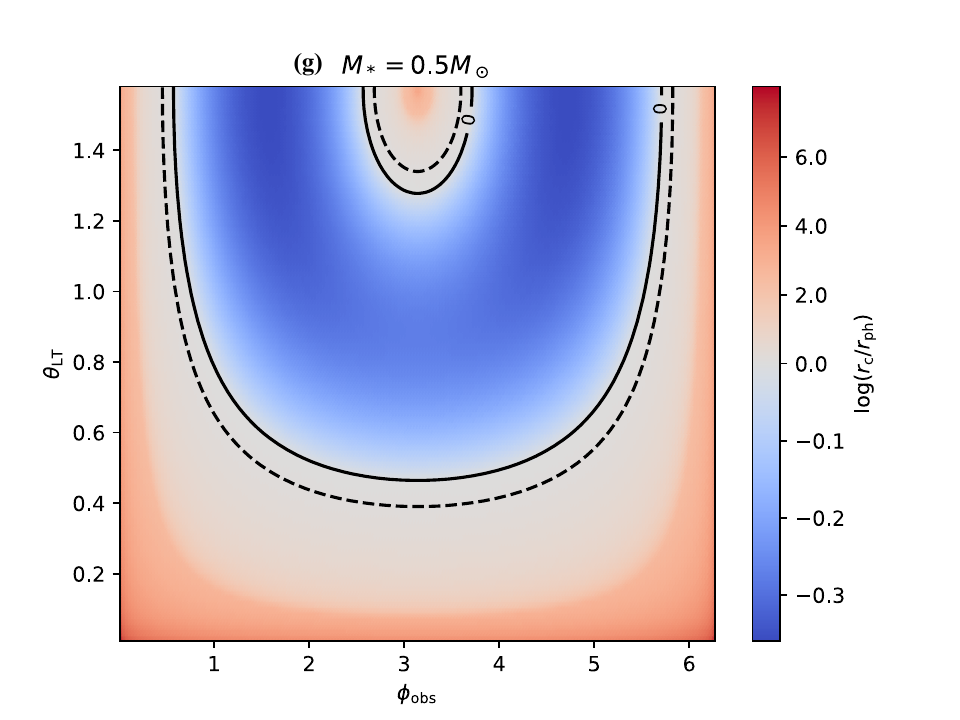}
\includegraphics [angle=0,scale=0.35] {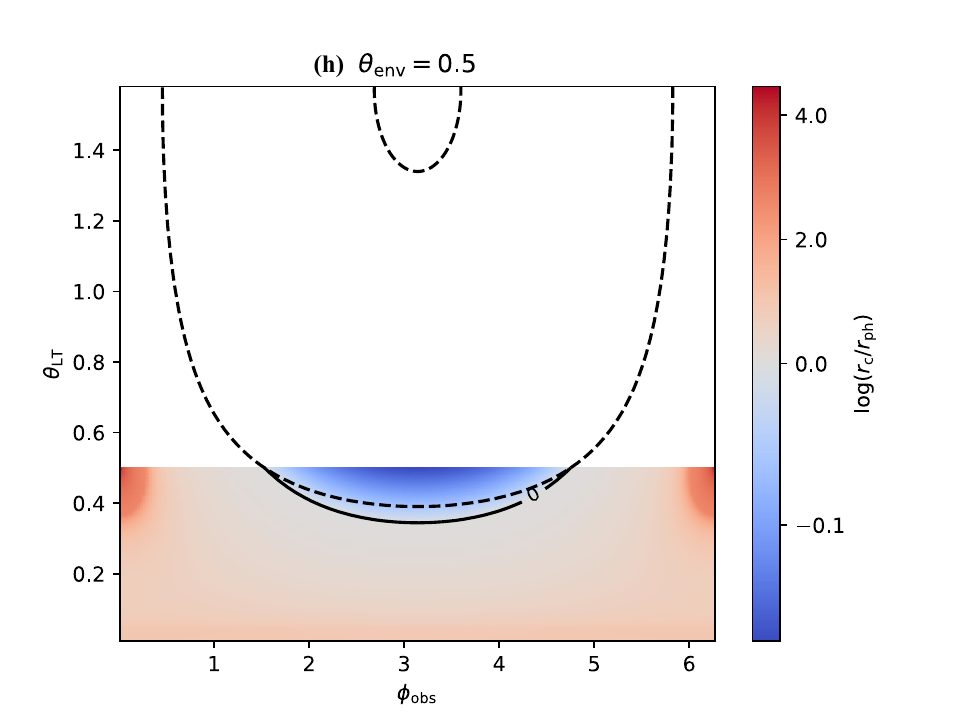}
\includegraphics [angle=0,scale=0.35] {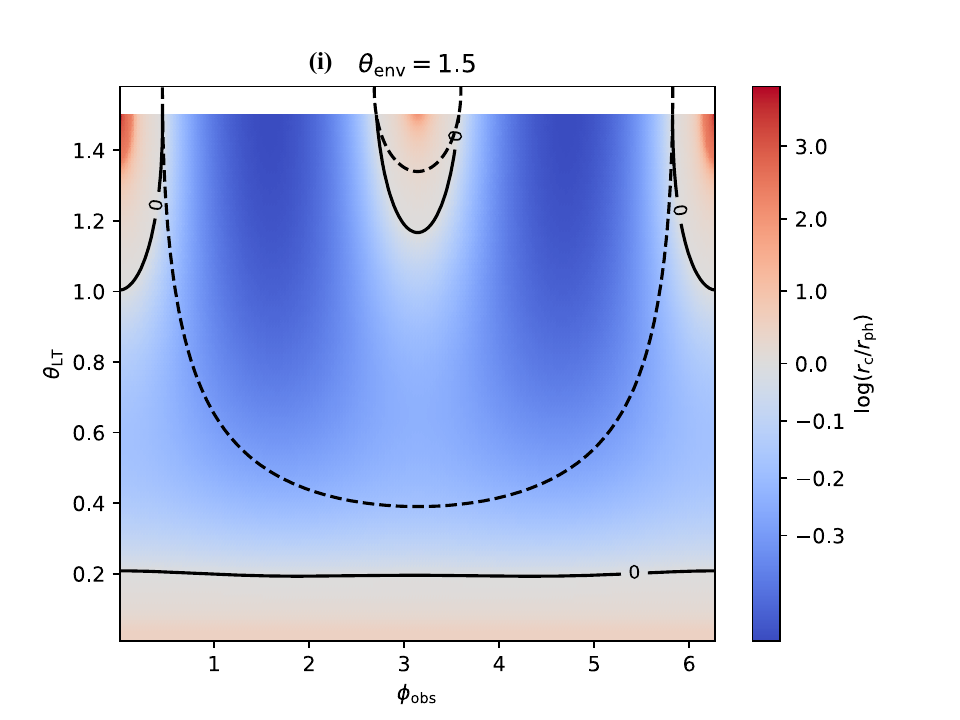}
\caption{
The logarithm of the ratio between the reverse-shock crossing radius and the photosphere of the envelope for the jet, $\log (r_{\rm c}/r_{\rm ph}$). Successful breakout jet requires $\log (r_{\rm c}/r_{\rm ph})>0$, as shown by the red area.
panel (a) is the fiducial case for $M_{\bullet}=10^6M_\odot$, $t=100~\rm day$, $\theta_{\rm j}=0.1$, $\epsilon=0.01$, $P_{\rm LT}=5~\rm day$, $M_*=M_\odot$ and $\theta_{\rm LT}=\theta_{\rm env}\in (0,\pi/2)$. In other panels, we only change one parameter for comparing with fiducial case;
Panel (b), $\epsilon=0.05$, panel (c), $M_{\bullet}=10^5M_\odot$, panel (d), $t=500~\rm day$; panel (e), $\theta_{\rm j}=0.4$; panel (f), $P_{\rm LT}=10~ \rm day$; panel (g), $M_*=0.5M_\odot$; panel (h), $\theta_{\rm LT}<\theta_{\rm env}=0.5$; panel (i), $\theta_{\rm LT}<\theta_{\rm env}=1.5$.
Besides, since larger $M_{\bullet}$ leads to a later onset of envelope evolution, we adopt the time of the peak $q$ to weaken the influence of the time.
The red and purple dashed lines are the results of \citet{Teboul-Metzger(2023)} and \citet{Lu-et-al.(2024)} for $\epsilon=0.01$ and $0.05$, respectively.
}
\label{fig:breakout condition}
\end{figure*}

The jet break out from the isotropic wind have been studied by \citet{Teboul-Metzger(2023)} and \citet{Lu-et-al.(2024)}. We show their results in the Fig. \ref{fig:breakout condition} using red (for \citet{Teboul-Metzger(2023)}) and purple (for \citet{Lu-et-al.(2024)}) dashed lines, respectively. The ZEBRA envelope contains a low density polar funnels where jet can freely escape, and is most likely inclined. We need to check whether the precessing jet can successfully break out from such inclined ZEBRA envelope. The parameter determining the geometry ($\theta_{\rm LT}$, $\theta_{\rm env}$, $\theta_{\rm j}$, $\phi_{\rm obs}$), the jet dynamics ($P_{\rm LT}$, $\epsilon$, $\Gamma_0$), envelope structure ($M_{\bullet}$ and $M_*$) will come into consideration.



In the calculation, we assume the polar angle of observer is equal to the precessing angle of the jet unless otherwise specified, $\theta_{\rm obs}=\theta_{\rm LT}$. We assume that the initial Lorentz factor of jet is $\Gamma_0=10$\footnote{One should note that \citet{Coughlin-Begelman(2014)} adopted the radiation powered jet. However, the numerical simulations by \citet{Sadowski-Narayan(2015)} and \citet{Sadowski-et-al.(2015)} found that the radiation alone cannot produce the high Lorentz factor of jet. The general relativistic radiation MHD simulations of accretion onto a SMBH during a TDE shows that a strong poloidal magnetic field and rapid BH spin are needed to produce the jet power in J1644 \citep{Curd-Narayan(2019)}. Therefore, we assume that the jet is produced by the influence of rapid SMBH spin and strong magnetic field, and adopt $\Gamma_0=10$ as a fiducial value in the calculation.}. 
We assume that the envelope properties remain nearly unchanged during the jet precession (The precessing period is usually much smaller than the duration of TDE).

Now, we investigate the effects of different parameter on jet breakout.
In Fig. \ref{fig:breakout condition}, we first define a fiducial case (panel a) with $M_{\bullet}=10^6M_\odot$, $t=100~\rm day$, $\theta_{\rm j}=0.1$, $\epsilon=0.01$, $P_{\rm LT}=5~\rm day$, $M_*=M_\odot$ and $\theta_{\rm LT}=\theta_{\rm env}\in (0,\pi/2)$. We then change the value of one parameter at each time and investigate its effect of this parameter on the jet breakout. For example, in panel b, we change the $\epsilon$ value as $\epsilon=0.05$, with other parameters as the fiducial case. In the same way, we change $M_{\bullet}=10^5M_\odot$ in panel c, $t=500~\rm day$ in panel d, $\theta_{\rm j}=0.4$ in panel e, $P_{\rm LT}=10~\rm day$ on panel f, and $M_*=0.5M_\odot$ in panel g. In panels h and i, we assume that $\theta_{\rm LT}<\theta_{\rm env}$, and take fix $\theta_{\rm env}=0.5$ in panel h and $\theta_{\rm env}=1.5$ in panel i. The blue shaded regions indicate parameter space that result in choked jets, i.e., $r_{\rm c}<r_{\rm ph}$. The solid black lines represent the boundary where jets marginally break out, i.e. $r_{\rm c}=r_{\rm ph}$. For comparison, in panels b-i, we also show the boundary lines of fiducial case with the dashed black lines. The results from \citet{Teboul-Metzger(2023)} and \citet{Lu-et-al.(2024)} are shown with red and purple dashed lines, respectively.
The main conclusions can be drawn as follow:
\begin{enumerate}
     \item In the fiducial case (panel a), the envelope density along its polar regions is significantly reduced, allowing the jets to escape freely from these funnels. Observers at $\theta_{\rm obs}=\theta_{\rm env}$ and $\phi_{\rm obs}$ (or $2\pi-\phi_{\rm obs})\lesssim 0.5$ can, therefore, detect a strong episodic jet. Away from these free zone, jets can only break out successfully at small precession angle cases. For observers at the choked zone, an off-axis jetted TDE is possibly available.  
     For $\theta_{\rm LT}\sim \pi/2$ and $\phi_{\rm obs}\sim \pi$, we can also observe the counter-jet emerging from the envelope.   
    \item As $\epsilon$ increases (panel b), the jet will become more powerful, and the breakout regions also increase. A smaller SMBH mass $M_{\bullet}$ (panel c) or a smaller stellar mass $M_*$ (panel g), resulting in a smaller ZEBRA photosphere size, allows a larger breakout zone. At late observation time (panel d), the envelope size increases, resulting in a smaller regions where jets can successfully break out. Variations in the jet opening angle (panel e) have little effect on the jet breakout, which is consistent with the results in \citet{Lu-et-al.(2024)}. As $P_{\rm LT}$ increases (panel f), the duty cycle also increases, and the choked zone become smaller. 
    \item For $\theta_{\rm LT} < \theta_{\rm env}$ (panels h and i), the jet remains visible at smaller precession angle cases or close to the polar funnels.
\end{enumerate}
The successfully breaking out jet will generate multi-band afterglow emission. If the jet is choked, thermal radiation in the optical, ultraviolet, or X-ray bands will be produced during the cocoon shock breakout \citep{Yuan-Lei(2024), Zhu-et-al.(2021)}, as shown in Fig. \ref{fig:ZEBRA_model}. 

\section{Discussion}\label{sec: discussion}

\subsection{Comparing with the model of the processing jet in the disk wind}
\citet{Teboul-Metzger(2023)} and \citet{Lu-et-al.(2024)} investigated the breakout conditions of precessing jets within the disk winds. Since a precessing accretion disk generates winds in different directions, these wind interact and eventually become isotropic. However, the density distribution of the ZEBRA envelope is not isotropic (consisting low density polar funnels), and the rotation axis is most likely inclined with respect to the SMBH spin axis. Therefore, we need to study the jet breakout at different precession angles and also azimuthal angles in ZEBRA case.

\citet{Teboul-Metzger(2023)} and \citet{Lu-et-al.(2024)} analyzed the influence of both the accretion-to-radiation energy conversion efficiency ($\epsilon$) and the jet precession angle ($\theta_{\rm LT}$) on the jet breakout. 
We present their results in Fig. \ref{fig:breakout condition}, with red dashed lines from \citet{Teboul-Metzger(2023)} and purple dashed lines from \citet{Lu-et-al.(2024)} for $\epsilon=0.01$ and $0.05$. Jet can only escape from the wind when it precesses with very small precession angles in these works. In our work, however, we find a new possible regions for the successful jet breakout at large precession angles, i.e., when the jet moves close to the polar regions of ZEBRA envelope. 
When the line of sight is far away from the polar axis, e.g., $2\lesssim \phi_{\rm obs} \lesssim 5$, our results are quite similar with that of \citet{Teboul-Metzger(2023)}. In \citet{Lu-et-al.(2024)}, they assume that most accreted material is expelled as disk wind, while in \citet{Teboul-Metzger(2023)}, only a fraction ($\sim 0.1$) is ejected as wind. This makes the significant discrepancies between their results (see the red and purple dashed lines in Fig. \ref{fig:breakout condition}).


\subsection{Jet/disk alignment and envelope collapse}
There are several mechanisms leading the jet/disk alignment as pointed out in \citet{Teboul-Metzger(2023)} and \citet{Lu-et-al.(2024)}: 1) The torque exerted by the fallback flow may rapidly damp the disc precession, and the timescale is in a few to tens of days for different accretion rate, viscosity, and SMBH mass \citep{Zanazzi-Lai(2019)}; 2) As the accretion rate decreases, the accretion disk gradually thins. During this process, the Bardeen-Petterson effect causes the disk's precession angle to progressively diminish, and the time scale of the alignment is $\sim 200$ days for $M_{\bullet}=10^6M_{\odot}$ \citep{Teboul-Metzger(2023)}; 3) For sufficiently powerful magnetized jets, the magnetic torques will align the jet/disk with the SMBH spin axis \citep{McKinney-et-al.(2013)}. The alignment timescale with this process is significantly shorter than other mechanisms \citep{Teboul-Metzger(2023)}.


Due the alignment, the initially choked precession jets have chance to escape at late time. As shown in \citet{Teboul-Metzger(2023)}, the delayed emergence of non-thermal radio afterglow can be explained with the alignment process induced by the thinning of a thick accretion disk. Such as AT2018hyz, after the discovery in the optical band, it showed no detectable radio emission for a long time, exhibiting only a flat thermal X-ray lightcurve \citep{Cendes-et-al.(2022), Gomez-et-al.(2020)}. However, after $\sim 972$ days, a bright radio afterglow suddenly emerged \citep{Cendes-et-al.(2022)},  which suggests the emergence of a non-relativistic jet at a very late stage, further indicating that such phenomena may originate from the late-time alignment of a precessing jet.
In addition, ASASSN-15oi \citep{Hajela-et-al.(2025)} and AT2019azh \citep{Goodwin-et-al.(2022)} also exhibit similar radiative characteristics and behaviors.

In our model, it is possible to detect a two bumps radio light curves, with the first bright one from the jet escaping from the envelope polar region, and the late one from the successful breaking out jet (but off-axis) due to the alignment.

Another point is the collapse of ZEBRA envelope as the accretion rate drops with time and falls below the Eddington limit \citep{Ferris2022}. \citet{Metzger2022} proposed a cooling envelope model, in which the envelope cools radiatively and undergoes Kelvin-Helmholtz contraction \citep{Metzger2022,Sarin2024}. The rising of SMBH accretion rate due to the collapse or contraction of ZEBRA envelope, may account for the delayed onset of thermal X-rays, and late-time radio flares \citep{Metzger2022}.


    

\subsection{Comparison with Observations}
\label{observations}
In the following calculations, we assume $\phi_{\rm obs}\sim 0$ for simplicity. Besides, the structure of ZEBRA envelope that we adopt is that in the fiducial case (see panel a in Fig.\ref{fig:breakout condition}), i,e,  $M_{\bullet}=10^6M_\odot$, $t=100~\rm day$ and $M_*=M_\odot$. We also take $\epsilon=0.01$ as in the fiducial case.

\begin{figure*}
\centering
\includegraphics [angle=0,scale=0.22] {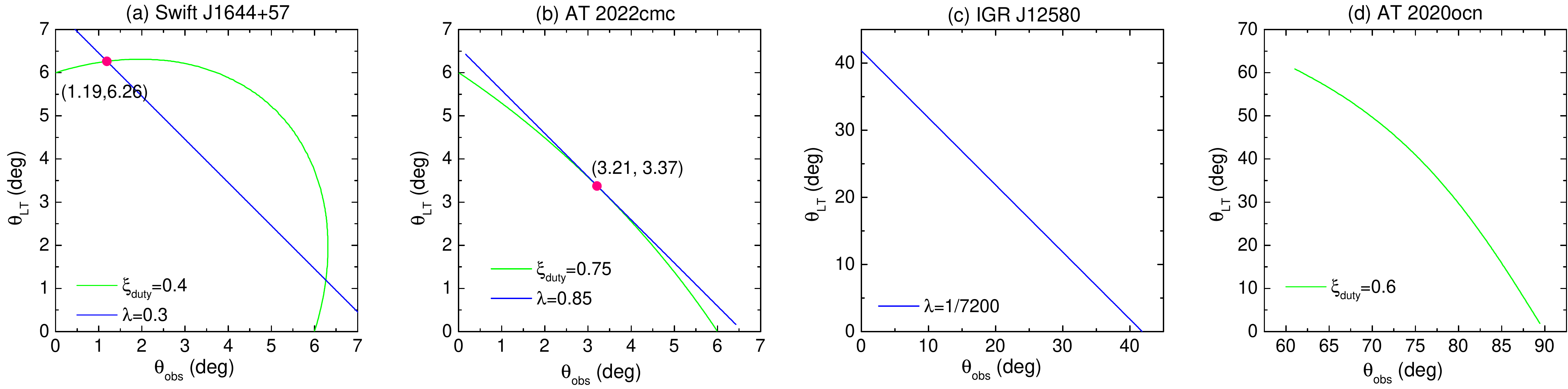}
\caption{
The estimation of $\theta_{\rm obs}$ and $\theta_{\rm LT}$ for Swift j1644+57 (panel a), AT 2022cmc (panel b), IGR J12580 (panel c) and AT 2020ocn (panel d) with the special $\xi_{\rm duty}$ or $\lambda$. The pink point in each panel represents the values of ($\theta_{\rm obs}$, $\theta_{\rm LT}$) that we adopt.
}
\label{fig:TDEFitting}
\end{figure*}

\begin{enumerate}
    \item Swift J1644+57 (J1644): a jetted TDE with precessing.\\
    The X-ray temporal evolution of J1644 exhibits periodical dips \citep{Bloom-et-al.(2011), Burrows-et-al.(2011), Levan-et-al.(2011)}.
    We adopt $P_{\rm LT}\sim 2.7~$day for J1644 \citep{Lei2013}. According to the Fig.\ref{fig:P_LT_spin}, one can estimate the spin of SMBH $a_\bullet\sim 0.75$.
    For jets with significant precession, the observed flux primarily depends on two factors: one is $\psi$, which is the angle between the line of sight and the jet axis, and the other is the jet Lorentz factors when the jet breaks out from the ZEBRA envelope. The evolution of $\psi$ leads to the on-axis and off-axis of the jet, i.e., the duty cycle $\xi_{\rm duty}$. For J1644, the duty cycle is roughly $\xi_{\rm duty}\sim 0.4$.    
    Since we observed strong emission and high Lorentz factor in J1644 jet, we assume the jet escape along the ZEBRA polar ($\theta_{\rm LT}\sim \theta_{\rm env}$).
    Besides, we assume the opening angle of the jet $\theta_{\rm j}\sim 6^\circ$ \citep {Lei2013}.  
    Another parameter is the peak-to-dip flux ratio $\lambda=F(\psi=\psi_{\rm max}, t)/F(\psi=0,Dt)$, where $\psi_{\rm max}$ is the maximum value of $\psi$. The observed flux is given by $F(\psi, t)=D^4F(\psi=0,Dt)$,
    where $D=\Gamma_{\rm on}(1-\beta_{\rm on})/\Gamma_{\rm off}(1-\beta_{\rm off}\cos\psi)$ is the ratio of the off-beam Doppler factor to the on-beam Doppler factor, $\Gamma_{\rm on}=1/\sqrt{1-\beta_{\rm on}^2}$ ($\Gamma_{\rm off}=1/\sqrt{1-\beta_{\rm off}^2}$) is the Lorentz factor for $\psi=0$ ($\psi>0$).
    In the ZEBRA envelope, the density distribution is not isotropic. When the jet precesses to different orientations, it sweeps through different density medium, resulting in different Lorentz factors when the jet breaks out. We adopt the method described in \citet{Huang-et-al.(1999)} to calculate the evolution of the jet Lorentz factor. 
    We use the empirical relationship between $\Gamma$ and $L_{\rm j,iso}$ to derive the jet initial Lorentz factor $\Gamma_{0}$ \citep{Wu-et-al.(2016)}, where $L_{\rm j,iso}=L_{\rm j}/(1-\cos\theta_{\rm j})=L_{\rm X,iso}/\epsilon_{\rm X}$ is the isotropic luminosity of the jet, $L_{\rm X,iso}$ is the isotropic radiation luminosity in X-ray and $\epsilon_{\rm X}\sim 0.1$ is the efficiency of jet luminosity to X-ray radiation. For J1644, $L_{\rm X,iso}\sim 10^{47}~\rm erg~s^{-1}$ and $\Gamma_{0}\sim 30$. We also use $\lambda\sim 0.3$ \citep{Lei2013}.
    The line of sight $\theta_{\rm obs}$ and precession angle $\theta_{\rm LT}$ should satisfy the constraints both $\xi_{\rm duty}$ and $\lambda$, see Fig.\ref{fig:TDEFitting}. 
    %
    One can find two different solutions, e.g., ($\theta_{\rm obs}=1.19^\circ$, $\theta_{\rm LT}=6.26^\circ$) and ($\theta_{\rm obs}=6.26^\circ$, $\theta_{\rm LT}=1.19^\circ$).
    J1644's jet is initially precessing but ceased after $\sim 30$ days due to the alignment with the SMBH spin \citep{Lei2013, Teboul-Metzger(2023)}. The jet emission is observed even after the alignment, we then expect a small $\theta_{\rm obs}$ (otherwise, the line of sight would be outside the jet cone after the alignment). Therefore, 
    for J1644, one can get $\theta_{\rm obs}=1.19^\circ$ and $\theta_{\rm LT}=6.26^\circ$. 
    During the late evolutionary phase of the ZEBRA, the accretion rate declines to the Eddington limit, leading to the shut-off of relativistic jet \citep{Coughlin-Begelman(2014)}, which corresponds to the steep decaying light curve in the late-time emission.
    The absence of detectable optical emission from J1644 is attributed to significant dust extinction within its host galaxy, which effectively suppresses the intrinsic UV/optical continuum radiation \citep{Burrows-et-al.(2011), Levan-et-al.(2011)}.

    \item AT2022cmc: a jetted TDE without precessing.\\
    Unlike J1644, no precession signatures have been observed in AT2022cmc. Observations show that its X-ray light curve rapidly transitions into a power-law decay phase during the early evolution \citep{Andreoni-et-al.(2022)}. 
    Within our theoretical framework, the absence of early jet precessing signatures arises from the nearly alignment of jet, envelope polar axis, and SMBH spin axis.
    Since we observe strong jet emission but without significant precession in this case, we assume the jet escapes along the ZEBRA polar axis ($\theta_{\rm LT}\sim \theta_{\rm env}$), but expect a small jet precesssing angle $\theta_{\rm LT}$ and observer angle $\theta_{\rm obs}$. 
    We adopt
    $\theta_{\rm j}\sim 6^\circ$, $L_{\rm x,iso}\sim 5\times 10^{46}~\rm erg~s^{-1}$, $\xi_{\rm duty}\sim 0.75$ and $\lambda\sim 0.85$ to estimate the values of $\theta_{\rm LT}$ and $\theta_{\rm obs}$ with the similar method for the J1644, as shown in the panel (b) of Fig.\ref{fig:TDEFitting}. We also use the $\Gamma- L_{\rm X,iso}$ empirical relation \citep{Wu-et-al.(2016)} to get $\Gamma_0\sim25$ for AT2022cmc. We find $\theta_{\rm obs}\sim 3.21^\circ$ and $\theta_{\rm LT}\sim 3.37^\circ$. 

    \item  IGR J12580 (J12580): a  TDE with an off-beam relativistic jet.\\
    IGR J12580 initially observed by Integral in had X-ray band \citep{Walter-et-al.(2011)}. Follow-up observations revealed that the X-ray light curve of J12580 maintained its peak luminosity for several tens of days before transitioning to $t^{-5/3}$ decay. While the overall evolutionary behavior resembles that of J1644, J12580 exhibits less variability in X-ray lightcurve \citep{Burrows-et-al.(2011), Zauderer-et-al.(2013)}.
    Approximately one year after its X-ray discovery, the corresponding radio afterglow of J12580 was detected by the Karl G. Jansky Very Large Array (JVLA) \citep{Irwin-et-al.(2015)}.
    We expect a successful jet from J12580, which escapes from the polar axis of ZEBRA envelope. We therefore assume $\theta_{\rm LT}\sim \theta_{\rm env}$. 
    The X-ray emission from the off-beam relativistic jet is strongly suppressed \citep{Lei2016,YuanQ2016}, and the radiation luminosity is $\sim 1/7200$ of J1644.
    In addition, we assume $\theta_{\rm j}\sim 8^\circ$ and $\Gamma_0\sim 5$ to estimate $\theta_{\rm obs}$ and $\theta_{\rm LT}$ for $\lambda=1/7200$. The result is shown in the panel (c) of Fig.\ref{fig:TDEFitting}. 
    Since only one constraint is applied, all points along the blue solid line are potential solutions. Since no significant precession observed, we expect the jet nearly aligns with the SMBH spin ($\theta_{\rm LT} \sim 0$), and the solution with large $\theta_{\rm obs}$ ($\sim 40^\circ$) will be consistent with \citet{Lei2016}. 
    The radio emission from the off-beam jet become visible when the jet is decelerated to non-relativistic \citep{Lei2016,YuanQ2016}. 

    \item ASASSN-14li: a TDE with the UV/radio emission and thermal X-ray component.\\
    ASASSN-14li was initially discovered through optical emission, followed by subsequent detections across X-ray, UV, and radio emission, and X-ray emission is dominated by thermal emission\citep{Miller-et-al.(2015), Holoien-et-al.(2016b)}. The existence of a weak jet is supported by the correlation between radio and X-ray flares \citep{Pasham2018}.
    In our model, the jet is weak \citep{Pasham2018}.
    Due to no significant preceesing signal in the UV band emission, we assume the jet is aligned with the SMBH spin and the line of sight is also along the SMBH spin. 

    
    \item AT2020ocn: a thermal X-ray TDE with QPO. \\
    AT2020ocn is an optical transient from the centre of a previously quiescent galaxy \citep{Gezari-et-al.(2020)}. Subsequently, the Neutron Star Interior Composition Explorer (NICER) started to observe and found that, unlike the optical and ultraviolet emission, the soft X-ray radiation exhibited clear quasi-periodic modulations with a period of $\sim 15$ days \citep{Pasham2024}. 
    The SMBH spin $a_\bullet\sim 0.25$ can be estimate from Fig.\ref{fig:P_LT_spin}.
    Moreover, the periodic X-ray emission shows two thermal components \citep[a cool and warm component, and the temperature of thermal components also modulates on the same time scale, see][]{Pasham2024}, indicating that the line of sight aligns with the envelope’s inclination axis ($\theta_{\rm obs}\sim \theta_{\rm env}$). In this configuration, the observer can detect not only the thermal radiation from the inner thick disk within the envelope, but also the thermal emission produced by the cocoon shock breakout after the precessing jet is choked.
    If the cocoon shock breakout occurs on the other side (relative to the line of sight) of the ZEBRA surface, it would be obscured by the envelope, leading to periodic variations due to the jet precession\footnote{Whether the cocoon shock breakout radiation at the envelope surface can be observed depends on the angle between the observer’s viewing direction ($\vec{r}_{\rm obs}$) and the normal of the surface element at the shock breakout region ($\vec{r}_{\rm co}$), only when $\vec{r}_{\rm obs}\cdot \vec{r}_{\rm co}>0$ (or the angle is less than $90^\circ$), the radiation can be observed.}.
    In our model framework, the envelope of AT2020ocn has a large inclination angle, and the jet precession angle is smaller than the envelope inclination.   
    We constrain the values of $\theta_{\rm obs}$ and $\theta_{\rm LT}$ for $\xi_{\rm duty}\sim 0.6$ in the panel (d) of Fig.\ref{fig:TDEFitting}. 
    Since only one constraint is applied, all points along the green solid line are potential solutions. Considering the significant precession expected from observations, we prefer a large $\theta_{\rm LT}$.   
    Adopting the ZEBRA envelope parameters in the fiducial case, 
    and assuming that the cocoon shock breakout is the Compton equilibrium between electrons and photons, one can calculate the cocoon breakout velocity $v_{\rm c,bo}\sim 0.1c$ with the observed temperature $T_{\rm c,bo}\sim 0.1~keV$ by the method in \citet{Zhu-et-al.(2021)} and \citet{Nakar-Sari(2010)}. In the calculation, we use $\theta_{\rm j}\sim 5^\circ$, the breakout position $r_{\rm c,bo}\sim 0.9 r_{\rm out}$ and density $\rho_{\rm c,bo}\sim 10^{-11}\rm g~cm^{-3}$ at this position.
    We can also rough estimate the value of the initial Lorentz factor ($\Gamma_0$) of the jet using the dynamical evolution equations for the jet and cocoon proposed by \citet{Huang-et-al.(1999)} and \citet{Zhang-et-al.(2024)} with the assumption of the jet choked position near the inner boundary of the envelope, and $\Gamma_0\sim 5$.
    Besides, similar modulations are
    not present in the optical bands \citep{Pasham2024}, which suggests that the optical radiation likely originates primarily from thermal emission at the envelope surface.


\end{enumerate}

\section{Conclusion}\label{conclusion}
The jet in TDEs is likely misaligned with the SMBH spin and precess due to the Lense-Thirring effect. We investigate the propagation of the precessing jet in a inclined ZEBRA envelope. We also study the effects of parameters ($P_{\rm LT}$, $\theta_{\rm LT}$, $\theta_{\rm env}$, $\theta_{\rm j}$, $\phi_{\rm obs}$, $\epsilon$, $r_{\rm out}$, $M_*$, and $M_{\bullet}$) on the jet propagation. Our conclusions are summarized as follows:

\begin{enumerate}

     \item The envelope density along its polar regions is significantly reduced, allowing the jets to escape freely from these funnels. For $\theta_{\rm LT} = \theta_{\rm env}$, observers with line of sight along these free zones can, therefore, detect a strong episodic jet, as the X-ray quasi-perodic dip shown in J1644. Away from these free zones, jets will be choked in most cases, and can only break out successfully at small precession angle. An off-axis jetted TDE, like J12580, is possible when looking along the choked zone.
    
    \item As $\epsilon$ increases, the jet will become more powerful, and the breakout regions also increase. A smaller SMBH mass $M_{\bullet}$ or a smaller stellar mass $M_*$, resulting in a smaller ZEBRA photosphere size, allows a larger breakout zone. Variations in the jet opening angle have little effect on the jet breakout, which is consistent with the results in \citet{Lu-et-al.(2024)}. As $P_{\rm LT}$ increases, the duty cycle also increases, and the choked zone become smaller. 

    \item At late time, the envelope size increases, resulting in a smaller regions where jets can successfully break out. 

    \item For $\theta_{\rm LT} < \theta_{\rm env}$, the jet remains visible at smaller precession angle cases or close to the polar funnels.

    \item The late brightening radio emission may come from the late-time alignment and breakout of the precessing jet, or the collapse or contraction of ZEBRA envelope. 

    \item Based on the distinct fates of precessing jets within the envelope, combined with the damping of jet precession and the influence of viewing perspectives, we can explain different observational phenomena in TDEs (see the discussions in Sec. \ref{sec: discussion}), e.g., J1644, IGR J12580, AT 2022cmc, ASASSN-14li, AT2020ocn and other similar events.

\end{enumerate}

\begin{acknowledgements}
We are very grateful to Xiangyu Wang, and Rongfeng Shen for their helpful discussions. This work is supported by the National Natural Science Foundation of China under grants 12473012 and 12533005, and the National Key R\&D Program of China (Nos. 2020YFC2201400, SQ2023YFC220007). W.H.Lei. acknowledges support by the science research grants from the China Manned Space Project with NO.CMS-CSST-2021-B11.
\end{acknowledgements}

\bibliography{export-bibtex}{}

\end{document}